# Compact arrangement for femtosecond laser induced generation of broadband hard x-ray pulses

C. Giles,[a,1] R. Celestre,[a] K.R. Tasca,[a] C.S.B. Dias,[a] R. Vescovi,[a] G. Faria,[a] G.F. Ferbonink,[b] and R.A. Nome[b,2]

[a]*Institute of Physics Gleb Wataghin, State University of Campinas, Campinas, SP, 13083-859, Brazil.*
[b]*Institute of Chemistry, State University of Campinas, Campinas, SP, 13083-970, Brazil.*

We present a simple apparatus for femtosecond laser induced generation of X-rays. The apparatus consists of a vacuum chamber containing an off-axis parabolic focusing mirror, a reel system, a debris protection setup, a quartz window for the incoming laser beam, and an X-ray window. Before entering the vacuum chamber, the femtosecond laser is expanded with an all reflective telescope design to minimize laser intensity losses and pulse broadening while allowing for focusing as well as peak intensity optimization. The laser pulse duration was characterized by second-harmonic generation frequency resolved optical gating. A high spatial resolution knife-edge technique was implemented to characterize the beam size at the focus of the X-ray generation apparatus. We have characterized x-ray spectra obtained with three different samples: titanium, iron:chromium alloy, and copper. In all three cases, the femtosecond laser generated X-rays give spectral lines consistent with literature reports. We present a rms amplitude analysis of the generated X-ray pulses, and provide an upper bound for the duration of the X-ray pulses.

## 1. INTRODUCTION

The stroboscopic principle of femtosecond spectroscopy requires spectrally broad pump and probe pulses in order to achieve atomic resolution in both space and time. Originally, ultrafast dynamics studies had been realized with femtosecond lasers in the visible spectral range. Broadband laser-induced creation of vibrational wave-packets leads to phase-space focusing of molecular structures. With the aid of molecular modelling of the corresponding energy landscapes, high spatial resolution is achieved [1]. Alternatively, sources producing light or matter waves with sub-angstrom wavelength are required to achieve real atomic resolution. Thus, one of the new key enabling technologies in the field of ultrafast structural dynamics is the development of femtosecond hard X-ray sources. Fortunately, in the past couple of decades we have witnessed remarkable progress on the generation of ultrashort hard X-ray pulses[2,3]. Within this context, several approaches have been developed for the production of femtosecond hard X-rays, ranging from table-top femtosecond X-ray plasma sources and electron guns to large facilities producing X-ray Free Electron Lasers. As a result, recent developments in time-resolved spectroscopy experiments employing such sources have allowed scientists to obtain detailed "molecular movies" of fundamental processes in physics, chemistry, and biology[4-10].

Among these sources, table-top fs X-rays are arguably the simplest to produce and have unique properties such as short pulse duration, small source size, and high brilliance[11,12]. Femtosecond laser-based production of X-rays in solid sample targets was discovered in 1991[4], and it employs a principle which later became known as the $K_\alpha$ X-ray source[13]. The mechanism relies on the generation of fast electrons by the extremely high electric field intensity of the femtosecond optical pulses, which eject core shell electrons off the target sample, thereby producing characteristic X-ray lines. The characteristic X-ray radiation originates from the transition of an outer-shell electron into a core hole in the target atoms, thus reproducing the same process employed in conventional X-ray tubes (with the characteristic X-rays followed by Bremsstrahlung). This type of source is called $K_\alpha$ source, since the X-rays are produced by fluorescence, thus emitting characteristic spectral lines in isotropic fashion over $4\pi$ radians[14].

Recently, we have reported our research on the development and application of fluorescence spectroscopy apparatus for measurements inside pressure chambers [15] as well as femtosecond pump-probe and transient grating spectroscopy [16], and femtosecond Raman-induced OKE spectroscopy [17,18]. In the present work, we report the design and construction of a broadband pulsed $K_\alpha$ X-ray source with spectral bandwidth ranging from 3 keV to 9 keV, with short pulse duration and photon flux of $10^{2-3}$ photons/pulse at 1 kHz.

## 2. DESIGN AND CONSTRUCTION

The first generation of fs plasma X-ray sources employed lasers with high peak energies (100 mJ/pulse) at low repetition rates (10Hz) [4-6]. Over time, the trend moved towards the use of kHz systems with pulse energies in the 1-10 mJ range focused down to spots $\leq$ 10 μm in diameter at the target [7-10,19]. In the present work, we have employed a regenerative chirped pulse amplifier (Coherent, Legend) with 1mJ pulse energy, 70 fs pulse duration, 800 nm center wavelength, and 1 kHz repetition rate. By focusing the beam down to 10 μm spot size, we estimate the focal intensity to be $1.8 \times 10^{16}$ W/cm$^2$, which is higher than the threshold for plasma formation and X-ray pulse generation by fs-laser excitation[20,21].

In order to focus the femtosecond laser pulse to a spot size of a few micrometers while maintaining its short 70 fs pulse duration and 1mJ pulse energy, we employed an all-reflective design comprised of a beam expansion telescope and parabolic focusing mirror. Thus, before entering the vacuum chamber used for X-ray generation, the femtosecond laser beam waist (1 cm) is increased by a factor of 6 using a pair of convex and concave curvature high energy dielectric mirrors (CVI) with anti-reflection coating (reflectivity >99%) and low wavefront distortion (λ/20). To further minimize wavefront aberrations that crucially affect efficiencies of both light focusing and nonlinear optical X-ray generation, the first (convex) telescope mirror is moved laterally by 5 mm (normal to the incident beam). Such alignment strategy obviates the need for off-axis tilting of the telescope mirrors thereby avoiding astigmatism and coma. The second curved (concave) mirror gives a final beam waist of 6 cm, which is slightly less than the open aperture of the focusing mirror. We employ an additional pair of flat mirrors (Newport) between the curved mirrors to minimize the overall telescope setup footprint.

---
[1] Deceased.
[2] Corresponding author: nome@iqm.unicamp.br

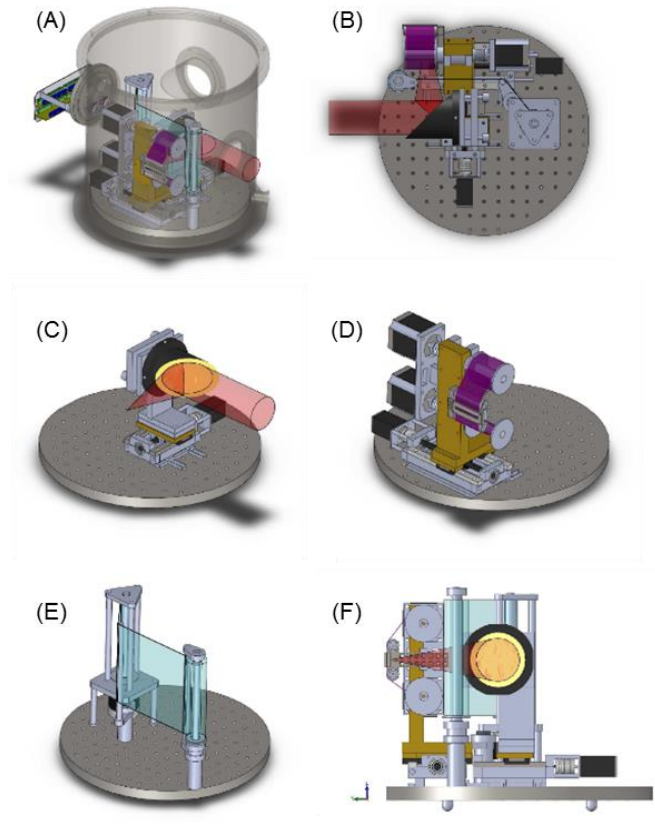

Figure 1. Design of femtosecond hard X-ray pulse generation source: (A): chamber view with all components employed; (B) and (C): focusing system; (D): target sample positioning system; (E): debris protection; (F): chamber side view.

Femtosecond-induced pulsed X-rays were generated in a vacuum chamber ($10^{-3}$ mbar) to increase their overall photon flux and avoid undesired plasma formation in open air. Figure 1 shows our vacuum chamber design, including opto-mechanical and motion control components employed in the present work. Specifically, the main components of the chamber are: focusing system, sample positioning reel system, and debris protection system. We have built the chamber shown in Figures 1 and 2 and used it to study femtosecond laser induced X-ray generation. Thus, after the beam expansion all-reflective telescope, the beam is sent through a sapphire window into the vacuum chamber (Figure 1A) for femtosecond laser-induced X-ray generation. Inside the chamber, a three-inch off-axis parabolic mirror (f = 150 mm, Janos Technology) is employed for optimum focusing efficiency at the sample position (Figures 1B and 1C).

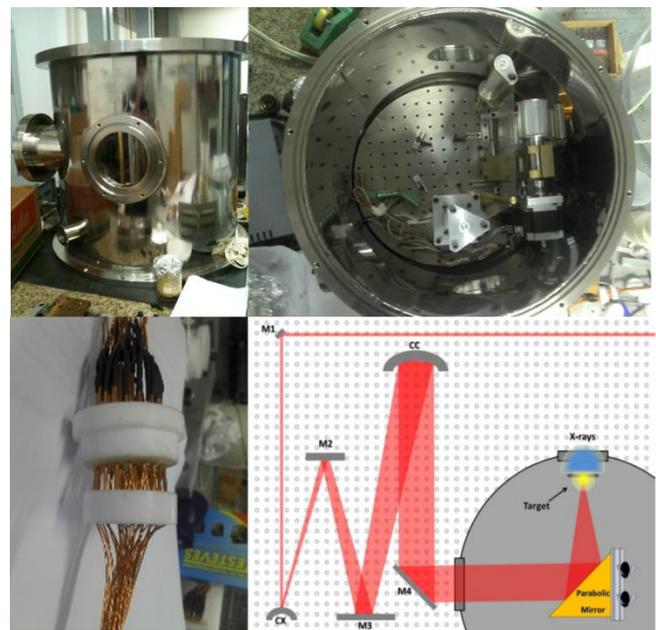

Figure 2. Photographs of the chamber employed for femtosecond hard X-ray pulse generation, showing the chamber windows (Top, Left), interior view with reel and debris protection systems (Top, Right), Be window (Bottom, Left), and optical layout (Bottom, Right).

In addition to achieving high peak intensities needed for fs X-ray pulse generation, the laser beam spot size at the focus also determines the X-ray target size. The target size, in turn, affects both X-ray transverse coherence length and overall sample/sample holder design. Thus, the target is placed as close as possible to the X-ray exit window to optimize the overall transverse coherence length, and we employ back illumination of the metallic target to further improve X-ray photon collection efficiency. Finally, the focusing mirror is placed on a three-axis motorized translation stage (Newport) such that the beam is focused at the target position.(Figures 1C-1F).

During the X-ray generation process, each femtosecond laser pulse incident on the solid target must interact with a new portion of the target sample, since the sample is destroyed upon multiphoton ionization interaction with the high peak intensity laser employed. Although its position is fixed along the laser beam propagation direction, the target must move in the transverse directions, besides being thin and stable. For this reason, the solid target consists of a metallic strip (Sandinox) mounted on a XY linear translation stage. This translation stage moves approximately 10 μm at every 1 ms to ensure that every femtosecond laser pulse interacts with a fresh portion of the target. Furthermore, to ensure that this X-ray source operates continuously for several hours, the moving sample strip must have a 50 m length and 50 mm width. With the XY translation stage, the sample is raster scanned such that the entire target sample material is used for X-ray generation. We implement single-shot X-ray pulse generation with the reel system shown in Figures 1D (design) and Figure 2 (photograph). Motion control of the parabolic mirror and the reel system is performed by in-house software

(*baremetal*) that controls a development board (NXP). An additional reel system rolls a kapton foil (Figures 1E and 1F) between the focusing mirror and the metallic sheet target. This setup protects the mirror as well as both input and exit optical and X-ray windows against debris produced when the femtosecond laser hits the target. Furthermore, since the entire opto-mechanical assembly is powered inside the vacuum chamber, electrical contacts were implemented using copper wires passing through a Teflon spacer (Figure 2). Finally, the femtosecond-laser induced X-ray pulses generated are sent through a Be window to an X-ray detector (Amptek, XR-100CR) with over 80% quantum efficiency in the 3-15 keV spectral range employed herein.

## 3. RESULTS AND DISCUSSION

We have performed spatial and temporal characterization of the femtosecond laser pulses. The 1 mJ laser beam was attenuated by 7 dB prior to performing these measurements. The pulse exiting the amplifier initially had a TEM00 beam shape with 1 cm waist as measured by a commercial beam profiler (Thorlabs). Spatial characterization of the focused beam, in turn, was performed with the knife-edge technique. Given the 1:6 expansion telescope and the parabolic mirror employed in our optical setup, measuring the laser beam spot size at the focus is technically challenging. Thus, we have employed a high-resolution linear translation stage (Newport GTS150) with 30 nm step size for beam waist characterization at the focus. With the knife-edge technique, we have verified that the beam employed has 10 μm spot size at the parabolic mirror focus. Thus, we have successfully achieved small beam size at the focus for efficient X-ray generation. Nonetheless, we note that beam waist in the orthogonal direction was not measured, and we assume optimum alignment of the parabolic mirror in our analysis of peak intensity achieved in the current setup.

Unlike conventional X-ray tubes, in the case of femtosecond-laser-based X-ray generation, X-rays are produced only during the light-matter interaction time, which is dictated by the laser pulse duration. The femtosecond laser pulses were characterized in the time-domain by second-harmonic generation cross-correlation frequency-resolved optical gating (SHG-XFROG) by placing the vacuum chamber window in one of the arms of the SHG-FROG apparatus. We have employed the beam exiting the amplifier as the reference pulse, which was also characterized SHG-FROG. Starting with transform-limited pulses, we place the vacuum chamber input optical window in the laser beam path and seek to minimize group velocity dispersion (GVD). Thus, we re-optimize the grating-to-grating separation in the regenerative amplifier pulse compressor arm. At this stage, we focus the beam with the parabolic mirror and maximize brightness of the white light continuum that is generated in air. Second, with the full femtosecond hard X-ray generation station operating, we repeat the optimization procedure just described except that we now maximize X-ray intensity. Finally, we measure spectrograms of the optimized femtosecond pulses with second-harmonic generation frequency resolved optical gating (see below). The FWHM pulse duration retrieved from our measured spectrograms was 110 fs, which is slightly longer than the reference pulse duration (70fs). The only transmissive optical components in our setup shown in Figures 1 and 2 are air (outside the vacuum chamber) and the 1 cm thick quartz window placed in the vacuum chamber input. Both quartz and air are transparent in the laser spectral window so we can neglect windowing/narrowing as the cause of pulse duration broadening.

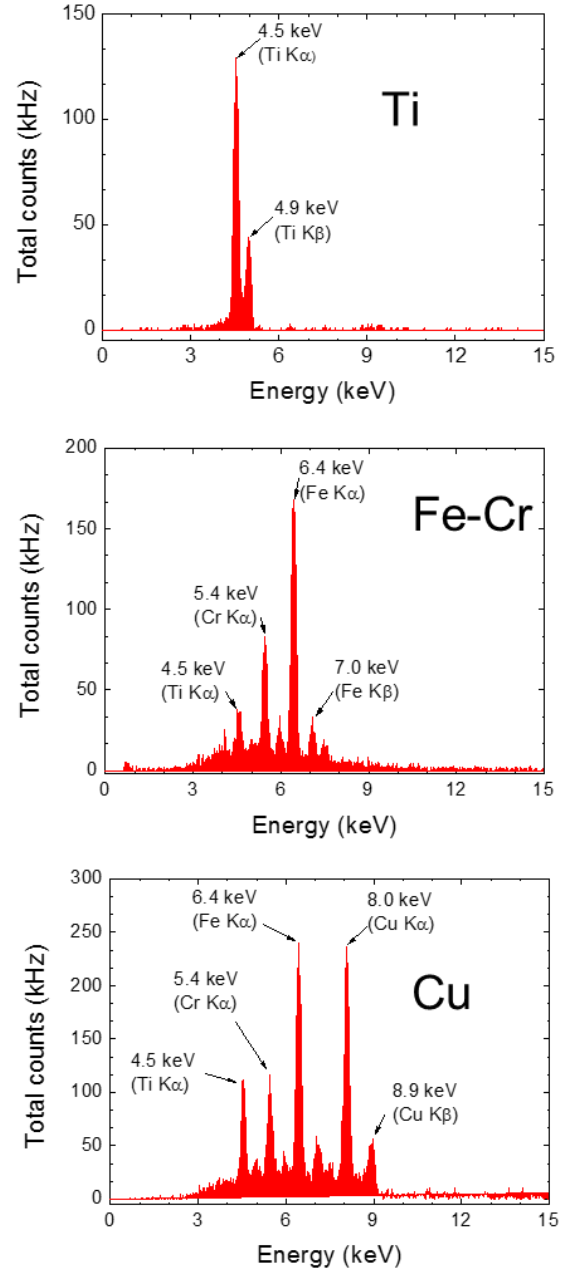

Figure 3. Femtosecond-laser induced X-ray Spectra plotted as counts per second (in Hertz) as a function of energy (in eV) (Top) Ti target; (Center) Ti, Cr, and Fe target, (Bottom): Ti, Cr, Fe, and Cu target. Energy resolution: 146 eV at 5.9 keV.

Given the pulse duration employed in the present work, the air GVD can also be safely neglected in the analysis of the temporal characterization of the laser pulses. Although we have compensated for the GVD imparted on the beam centered at 800 nm by the quartz window, we did not compensate higher-order dispersion that led to a small pulse broadening relative to the reference pulse.

Overall, the combination of all-reflective beam expansion telescope and parabolic focusing mirror helped us achieve high peak intensities for two main reasons. First, we were able to focus the femtosecond laser beam to a small spot size without employing lenses and objectives. In general, transmissive optics can lead to beam attenuation, wavefront distortion, and chromatic aberration, although we note that such effects can in principle be minimized by using high quality, AR coated, achromatic doublet lenses. Nonetheless, transmissive optics still result in dispersion stretching the pulse in time plus the accumulated nonlinearity/high B-integral also leads to wavefront aberration and self-focusing. Second, the reflective design (except for the vacuum chamber window) allowed us to keep the FWHM pulse duration of 100 fs without resorting to additional prism or grating pulse compressors, which are inefficient and would not separately correct for third-order dispersion. Thus, the setup shown in Figures 1 and 2 optimizes the peak intensity by minimizing both pulse duration and beam waist while maximizing the pulse energy.

The setup presented in this work is versatile enough to explore different materials as targets. Indeed, we have investigated X-ray generation using different target sheets composed by pure elements (titanium) as well as alloys. Furthermore, by combining target sheets, we were able to generate broadband hard X-ray pulses as shown in Figure 3. Figure 3 (Top) shows our femtosecond laser-induced X-ray spectrum obtained with the setup shown in Figures 1 and 2, and using Ti as the target. The spectrum shown in Figure 3(Top) exhibits two peaks at 4.5 keV and 4.9 keV, which respectively correspond to the energies of the $K_\alpha$ and $K_\beta$ emission lines of Ti. Likewise, Figures 3(Center) and 3(Bottom) show our femtosecond laser-induced X-ray spectra of hybrid targets respectively containing Ti and Fe:Cr alloy (Figure 3, Center) and Ti, Cu, and Fe:Cr alloy (Figure 3, Bottom). The spectral assignments indicated in Figures 3(Center) and 3(Bottom) show the $K_\alpha$ and $K_\beta$ emission lines of Ti, Fe, Cr, and Cu. Thus, we have successfully used the regenerative amplifier to produce hard X-rays in the 3 - 9 keV range ($\lambda$ = 0.14 - 0.41 nm) employing the setup shown in Figures 1 and 2. Moreover, in going from Figure 3(Top) to Figure 3(Bottom), we note an overall X-ray fluorescence lineshape broadening. The inhomogeneously broadened lineshape shown in Figure 3(Bottom) results in emitted X-rays with similar transverse coherence length as the X-rays emitted from the pure Ti target (Figure 3, Top). Nonetheless, the broader emission of the hybrid target may find useful applications in both time and frequency domain experiments.

Figure 4 shows a sample X-ray pulse train produced by femtosecond laser-induced X-ray pulse generation. Electronic time-gating was employed whereby the X-ray detector was triggered by the regenerative amplifier clock output, which was sent through a digital delay generator (Stanford Research Systems) for synchronization of the laser and X-ray pulses. As shown in Figure 3, we have produced X-rays pulses at 1 kHz repetition rates.

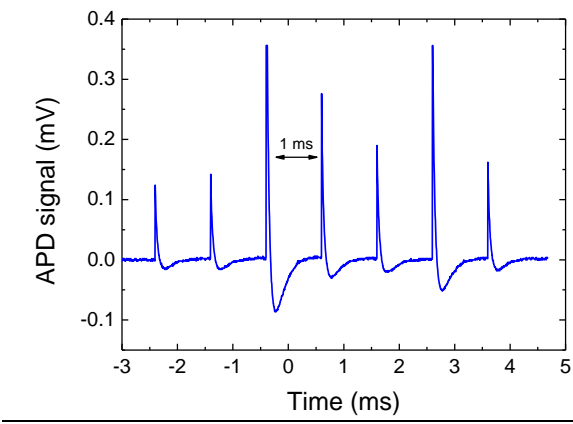

Figure 4. X-ray pulse train generated with the setup shown in Figures 1 and 2.

Analysis of the signal trace with a fast oscilloscope (Tektronix) shows that each pulse decays on the same timescale as the photodetector response. Given that the X-ray fluorescence we measure is a convolution of instrument response function with the material response, Figure 4 shows that we have generated X-ray pulses with pulse duration shorter than 5 ns.

In conclusion, we present a simple apparatus for femtosecond laser induced generation of X-rays. The apparatus consists of a vacuum chamber containing a parabolic focusing mirror, a reel system, a debris protection setup, a quartz window for the incoming laser beam, and an X-ray window for the generated X-ray beams. The femtosecond laser is expanded with an all reflective telescope design to minimize laser intensity losses and pulse broadening while allowing for focusing and peak intensity optimization. The laser pulse duration was characterized by second-harmonic generation frequency resolved optical gating. A high spatial resolution knife-edge technique was implemented to characterize the beam size at the focus of the X-ray generation apparatus. We have characterized x-ray spectra obtained with three different samples: titanium, iron:chromium alloy, and copper. In all three cases, the femtosecond laser generated X-rays give spectral lines consistent with literature reports. Finally, we present a rms amplitude analysis of the generated X-ray pulses, and we provide an upper bound limit for the duration of the X-ray pulses.

**Acknowledgment**. We hereby honor professor Carlos Giles (1964-2016) who had envisioned the development of integrated pump-probe laser/X-ray technologies for ultrafast structural dynamics measurements both table-top (present work) and in synchroton facilities (Brazilian Synchrotron Light Laboratory). We thank the Ultrafast Spectroscopy Laboratory of Prof. C. H. Brito-Cruz for the use of their femtosecond laser. We also thank Prof. Eckhart Förster, Institute for Optics and Quantum Electronics, Friedrich-Schiller-University Jena for advice throughout the research.